\documentclass[aps,pra,twocolumn,showpacs,groupedaddress,letterpaper,nofootinbib]{revtex4-1}

\usepackage{graphicx}
\usepackage{mdframed}
\usepackage{framed}
\usepackage{indentfirst}
\usepackage{natbib}
\usepackage{amsmath,amssymb}
\usepackage{subfigure}
\usepackage{enumitem}

\begin{document}

\title{Student Difficulties with Boundary Conditions in Electrodynamics}


\author{Qing X. Ryan}
\affiliation{Department of Physics, University of Colorado, 390 UCB, Boulder, CO 80309}

\author{Steven J. Pollock}
\affiliation{Department of Physics, University of Colorado, 390 UCB, Boulder, CO 80309}

\author{Bethany R. Wilcox}
\affiliation{Department of Physics, University of Colorado, 390 UCB, Boulder, CO 80309}

\begin{abstract}
Boundary conditions (BCs) are considered as an important topic that advanced physics undergraduates are expected to understand and apply. We report findings from an investigation of student difficulties using boundary conditions (BCs) in electrodynamics. Our data sources include student responses to traditional exam questions, conceptual survey questions, and think-aloud interviews. The analysis was guided by an analytical framework that characterizes how students activate, construct, execute, and reflect on boundary conditions. Common student difficulties include: activating boundary conditions in appropriate contexts; constructing a complex expression for the E\&M waves; mathematically simplifying complex exponentials and checking if the reflection and transmission coefficient are physical. We also present potential pedagogical implications based on our observations.
\end{abstract}
\maketitle

\section{\label{sec:intro}Introduction}

Student difficulties at both the introductory and upper-division level have been broadly investigated by the PER community \cite{Meltzer2012}. Upper division problem solving involves more complicated math and more sophisticated physics topics. A growing body of research suggests that upper-division students continue to struggle with problem-solving in these advanced physics topics \cite{Wilcox2013, pepper2012}. Some physics topics appear several times in different contexts across the advanced undergraduate physics curriculum \cite{Manogue2001}; one such topic is boundary conditions (hereafter BCs). BCs are used in a variety of contexts including classical mechanics, quantum mechanics, and throughout E\&M. Because of the importance and difficulty of this topic, we present a preliminary investigation of student difficulties by focusing on the use of BCs\footnote{In the case of no free charge/current at the boundary, boundary conditions are: $\vec E^\parallel_1=\vec E^\parallel_2$, $ \epsilon_1E^\perp_1= \epsilon_2E^\perp_2$, $B^\perp_1=B^\perp_2$, $ \vec B^\parallel_1 / \mu_1=  \vec B^\parallel_2 / \mu_2$} in electrodynamics.

Problem solving is a complicated process, hence an organizational structure is often helpful in characterizing the nature of student difficulties and making sense of what they are struggling with \cite{Wilcox2014}. The ACER framework \cite{Wilcox2013} is such an analytical tool that characterizes student difficulties with upper-division problem solving by organizing the problem-solving process into four general components: \textit{Activation} of the tools, \textit{Construction} of the models, \textit{Execution} of the mathematics, and \textit{Reflection} on the results. These components appear consistently in expert problem solving \cite{Wilcox2014,Wilcox2013} and are explicitly based on a resources view on the nature of learning \cite{Hammer2000, Wilcox2013}. Since the particulars of using mathematical and physical tools to solve upper-division physics problems are highly context-dependent, ACER is designed to be operationalized for specific physics topics. Operationalization involves a content expert working through problems that exploit the targeted tool while carefully documenting their steps. This outline is then refined based on analysis of student work \cite{Wilcox2013}.

In this paper, we present an application of the ACER framework to students' use of BCs in electrodynamics. We summarize a few common student difficulties and discuss some implications for teaching.

\section{\label{sec:methods}Methods}

We collected student work from three sources: traditional midterm exam solutions (two semesters) from CU (N=128) \footnote{Since we are interested in the frequency of certain mistakes, for exam data, we are reporting the number of solutions instead of the number of students.}, the \textit{Colorado UppeR-division ElectrodyNamics Test} (CURrENT: a conceptual assessment given to many students in electrodynamics) \cite{Baily2012,Ryan2014} (N=278 from 6 different institutions), and two sets of think-aloud interviews (N=11) with CU students. 

Exams were analyzed by coding mistakes that appeared at each element of the operationalized framework. Exam questions mostly targeted Construction and Execution, and did not provide a complete picture of the Activation or Reflection components. To address all aspects of the ACER framework, we relied further on data from CURrENT and interviews. Another goal of the interviews was to further explore the nature of preliminary difficulties identified in the exams.

The electrodynamics course at CU (E\&M II: Griffiths \cite{Griffiths} Ch.7-12) is the second semester of the electricity and magnetism sequence. The student population is composed of physics, astrophysics, and engineering physics majors with a typical class size of 30-60 students. At CU, E\&M II is often taught with varying degrees of active engagement through the use of research-based teaching practices, such as peer instruction and in-class tutorials \cite{Chasteen2012}.

\section{\label{sec:results}ACER OPERATIONALIZATION}
\begin{figure}
\includegraphics[scale=0.5]{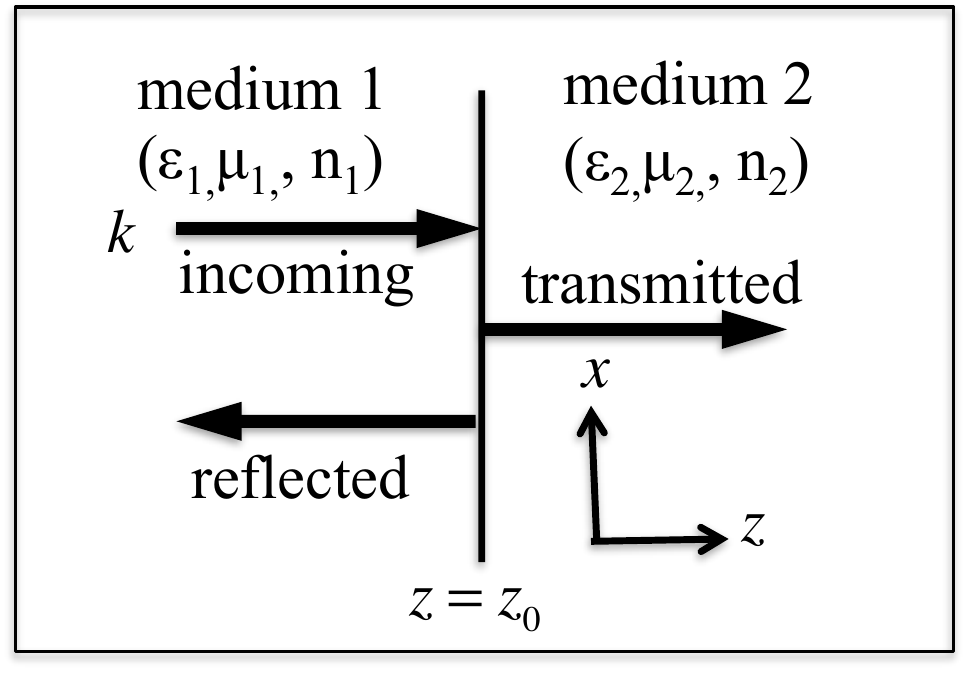}
\caption{Setup of a typical BCs problem in the context of electromagnetic waves}\label{fig:BCprob}
\end{figure}

We operationalized ACER for the use of BCs primarily in solving electromagnetic wave problems (Fig.\ \ref{fig:BCprob}). In this section, we provide a skeletal summary of the operationalized framework.

{\bf Activation of the tool}  involves identifying BCs as a relevant physical tool. BCs are likely to be activated in two situations:
\vspace{-2mm}
\begin{enumerate}[label=A\arabic*:, align=left] \itemsep-2pt
  \item Use BCs when being explicitly asked to (this would be considered as bypassing Activation).
  \item Use BCs when being presented a physical situation involving two media and asked for the relationship between physical quantities across the boundary.
\end{enumerate}
\vspace{-2mm}

{\bf Construction of the model}  involves mapping boundary conditions onto the specific physical situation/system. We operationalized Construction into three elements. The numbering of these elements is only for labeling purposes and does not necessarily indicate the order of the problem solving process. 
\vspace{-2mm}
\begin{enumerate}[label=C\arabic*:, align=left] \itemsep-2pt
  \item Write mathematical expressions of the complex fields for the incoming, reflected, and transmitted waves (hereafter denoted by the subscripts: in, refl, and trans).
  \item Select the appropriate BCs to be used with the corresponding components of the fields. 
  \item Set up equations at the boundary. This includes superposing the fields of the incoming and reflected waves, as well as applying the equations at the boundary.
\end{enumerate}
\vspace{-2mm}

{\bf Execution of the mathematics} involves performing mathematical steps to reduce the result to a form that can be interpreted. 
\vspace{-2mm}
\begin{enumerate}[label=E\arabic*:, align=left] \itemsep-2pt
   \item Algebraic manipulation to simplify the expression.
\end{enumerate}
\vspace{-2mm}

{\bf Reflection on the results} involves reflecting and evaluating results to gain physical insight and ensure consistency--a practice common among content experts. There are two common ways to reflect on BCs problems: 
\vspace{-6mm}
\begin{enumerate}[label=R\arabic*:, align=left] \itemsep-2pt
  \item Check the units of the relevant quantities.
  \item Check limiting behaviors. (e.g., the reflection coefficient R and transmission coefficient T must satisfy these conditions: $0 \leq R \leq 1$, $0 \leq T \leq 1$, $R+T=1$).
\end{enumerate}
\vspace{-2mm}

\section{\label{sec:discussion}RESULTS}
Using the operationalized ACER framework, we identified several common student difficulties. To organize the presentation of these difficulties, we group them according to the ACER components.

{\bf Activation of the tool:} Most exam questions in the context of electromagnetic (E\&M) waves bypass Activation by giving an explicit prompt (A1). Not only do these questions tell students to use BCs but also list the BCs (see footnote 1) explicitly for students. Therefore, in think-aloud interviews we investigated Activation by observing difficulties with Activation when BCs are not explicitly provided. 

In the interviews, we gave an E\&M wave problem to three students and asked them to use BCs but did not provide the equations.  Successful activation of BCs involves writing down $\vec E^\parallel_1=\vec E^\parallel_2$. One student was able to derive the BCs from Maxwell’s equations, but two students did not write down the BCs but rather, jumped straight to this result (with a sign error on $\tilde{E}_{0R}$): $\tilde {E}_{0I}=\tilde {E}_{0R}+\tilde {E}_{0T}$ (tilde means a complex quantity). The reasoning they provided to justify their answer looks like this: ``\textit{You can't just come away with more waves than you had coming in. Some of it will go through, some of it will go back. It has to add up, kind of like the conservation}." Their reasoning showed that they were not using BCs but rather, activated a different physics idea (an idea of conservation).

\begin{figure}
\includegraphics[scale=0.8]{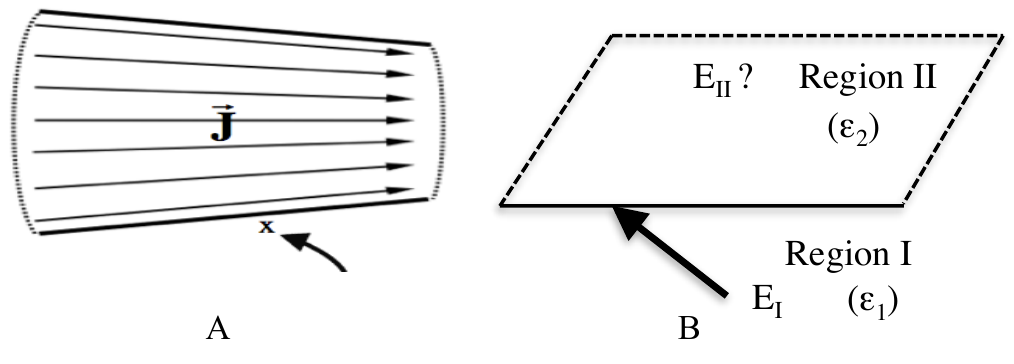}
\caption{A. Question on the CURrENT: A steady current flows in a long wire that has a uniform conductivity $\sigma$.  Is the electric field \underline{just} \underline{outside} the surface of the wire (e.g., at the point ``x" shown in the diagram) zero or non-zero? B. Scaffolding question: $ \epsilon_1= 2\epsilon_2$, $E_I$ in region 1 was shown in the picture. Use boundary conditions ($\vec E^\parallel_1=\vec E^\parallel_2$, $ \epsilon_1E^\perp_1= \epsilon_2E^\perp_2$) to draw an arrow that represents $E_{II}$ on the other side.}\label{fig:currentquestion}
\end{figure}

In addition to investigating student difficulties with Activation in the context of E\&M waves, we expanded our investigation by shifting to a different context. One question on the conceptual test (CURrENT \cite{Ryan2014}) asks if the E field just outside of a wire carrying steady current is zero or non-zero (Fig.\ \ref{fig:currentquestion}A) (this falls into ACER category A2). It also asks students to provide reasoning. According to BCs, the E field just outside of the wire must be non-zero. Only 30\% of the students (83 out of 278) explicitly used BCs, the rest of the students failed to Activate BCs on their own in this context.

We further explored the nature of student difficulties with the CURrENT question by conducting think-aloud interviews (N=6). Even after explicit or implicit cues from the interviewer, students still had trouble activating BCs. Different levels of cues were provided by the interviewer (cues were provided to 5 out of the 6 students): a) providing the Griffith's book\cite{Griffiths} and asked students to review the chapter on electromagnetic waves before the interview (1 student); b) providing a scaffolding question (Fig.\ \ref{fig:currentquestion}B) before giving the CURrENT question (3 students); c) asking students to redo this question once they solved an E\&M wave problem where BCs was activated (1 student).  

Two interviewees triggered by cues a and b respectively activated BCs. The remaining students were asked by the interviewer explicitly to use BCs once they failed to activate on their own. Even after being asked to use BCs, two students still didn't think that BCs were applicable. One student said: ``\textit{but that is light, this is not light}". Another student used BCs after being asked and correctly indicated the E is nonzero just outside of the wire but still was not satisfied with the answer: ``\textit{my answer is correct if the boundary conditions are true. I don't know if these boundary conditions were universal or only work for specific conditions.}"

We found that, in the absence of direct activation (A1), many students failed to activate BCs on their own. In the context of E\&M waves, some students still had trouble activating the correct BCs equations when they were not explicitly provided. When shifting to a different context (wire with steady current), students didn't think BCs were applicable.

{\bf Construction of the model:} When constructing a complex expression for the E and/or B field (e.g. $E_{0,in}exp{[}i(\vec{k} \cdot \vec{r}-\omega t){]}\hat{x}$) (element C1), errors commonly occur when simplifying $\vec {k} \cdot \vec{r}$ ($\vec{k}$ is the wave vector, $\vec{r}$ is the position vector). In the case of oblique incidence, the dot product becomes $k_xx+k_yy+k_zz$. Half of the solutions (18 out of 36) did not even consider $\vec {k } \cdot \vec{r}$. Out of those who considered the dot product, only 4 worked it out correctly. Various mistakes were made in the rest of the solutions: replacing $\vec {k} \cdot \vec{r}$ with $kz$, $kr$, $krsin\theta$ or $krcos\theta$. In the case of normal incidence where $\vec{k}$ is propagating in the z direction (Fig.\ \ref{fig:BCprob}), this dot product for the reflected and transmitted wave simplifies into ${-k_{refl}z}$ and $k_{trans}z$ with $k_{refl}$ and $k_{trans}$ being the magnitude of the wave vector in the two media. Working out the dot product for normal incidence is much easier and students performed better. However, almost a fifth (18\%, 23 out of 128) of the solutions missed the negative sign for $k_{refl}$, and more than a tenth (13\%, 16 out of 128) of the solutions did't differentiate between $k_{refl}$  and $k_{trans}$.

For normal incidence, students also need to recognize that E is parallel to the boundary and select the corresponding BCs (element C2).  About a tenth (11\%, 14 out of 128) of the solutions chose the wrong BCs equation. In the case of oblique incidence, students need to determine the correct components of E to be used with the corresponding boundary conditions (element C2). Over a third (33\%, 12 out of 36) of the solutions had the wrong components (e.g., use $sin$ instead of $cos$). There was not enough evidence from the written solutions to tell if students did not know what parallel or perpendicular means or simply made a trigonometry error. However, student interviews (N=4) on the same question suggest that the latter is unlikely to be the dominant reason.

Since there exist both incoming and reflected waves in medium 1 (Fig.\ \ref{fig:BCprob}), $\vec {E}$ in medium 1 can be written as $\vec{E}_{in}+\vec{E}_{refl}$ (element C3). Students need to replace $\vec {E}_{in}$  and $\vec {E}_{refl}$ with the complex expressions obtained in the previous step (C1). However,  in over a third (36\%, 46/128) of the solutions, E field in medium 1 was replaced with the sum of only the amplitudes of the incoming and reflected E field ($E_{0,in}+E_{0,refl}$). The exponential terms were lost in these solutions despite the fact that complex forms of the incoming and reflected E field were previously obtained. This strategy results in a correct expression only when the boundary is located at z=0 (Fig.\ \ref{fig:BCprob}) for normal incidence. 

It is hard from the limited work provided in exams to determine why students make such omission errors. However, the spontaneous comments offered by three students on their exam might shed some light on the reasoning. When ignoring the exponential terms and keeping only the amplitude of the E field, students arrived at the result: $|{E_{0,in}}|=|{E_{0,refl}}|+|{E_{0,trans}}|$. They justified their answer like this:``\textit{This makes sense since the incident wave gets partially reflected and partially transmitted}". This justification is similar to the conservation reasoning provided by the students interviewed (see previous section of ``Activation of the tool"). One might hypothesize that students omitted this piece of the ACER framework (C3) because of interference due to this conservation idea.

{\bf Execution of mathematics:} The most common mistake observed in execution (element E1) is the inappropriate cancellation of all exponential terms. When the boundary is located at z=0 (Fig.\ \ref{fig:BCprob}), exponential terms go to 1, so they can be dropped. However, students still cancelled the exponentials when the boundary was located at z=d. Over a third of the solutions (36\%, 32 out of 90) canceled the exponential terms inappropriately. The cancellation was coded as an Execution error in ACER; however, an execution error does not necessarily mean a math mistake. In other words, it doesn't indicate students would make such cancellation mistake in a pure math context. In order to further investigate the nature of this mistake, we conducted think-aloud interviews with a similar expression written in a pure math context. Symbols that represent physical quantities were replaced with arbitrary math variables (wave vector k, speed of light c, time t were replaced with d, a, b). None of the interview students made the same cancellation mistake. Interestingly, 3 (out of 4) students still made the connection to the wave context spontenously: “\textit{this is similar to the E\&M waves problems}”, “\textit{I remember you can always match the coefficients in those boundary condition problems}”. These interviews indicate that students’ math execution can be affected by the physics context. 

Even though students in this pure math context did not make the same cancellation mistake, they still struggled with complex exponential calculations. Two students made major math errors with complex exponentials: (e.g., $e^{A+B}=e^A+e^B$ and $e^{-A}=e^{-1}e^A$). 

{\bf Reflection on the results:} Evaluating a solution by checking units and limiting behaviors is an important skill that physicists value\cite{warren2010}. Our regular exam questions on BCs did not access Reflection explicitly and we saw almost no spontaneous reflection in the written work. Therefore we constructed an interview question to further investigate how students reflect. We presented students with incorrect but plausible expressions of the reflection and transmission coefficients R and T in a novel situation. Students were asked to examine if these expressions make sense without going through the calculation.  Three out of the 4 students interviewed were able to spontaneously point out that the T provided was not unitless (R1).

Compared to unit analysis, students had more difficulty with checking limits. For example, students had trouble identifying the independent variable (incident angle $\theta_I$). The interviewer had to provide a hint to direct them (3 out of 4) to think about the limits of R and T when $\theta_I$ varies from 0 to 90$^{\circ}$. The expressions of R and T given became negative or infinity when taking extreme limits of $\theta_I$. Only one student noticed the unphysical limits. One student obtained T\textless 0 and didn't note that this is unphysical. Two students talked about taking the lower limit of the index of refraction to 0. Students from the interviews struggled with identifying independent variables,  taking the limits of R and T, and recognizing  unphysical limits.

\section{\label{sec:discussion}CONCLUSIONS AND DISCUSSION}

When solving BCs problems, we found our students have difficulties in: (1) activating BCs in a context outside of E\&M waves, (2) constructing the complex expression for E and/or B field of the incoming, reflected and transmitted waves, (3) conducting and simplifying complex exponential math, (4) evaluating the reasonableness of the reflected and transmitted coefficient (R\&T) by checking if they have physical limits. 

These findings have several potential implications for teaching and assessing the use of BCs in electrodynamics and even across subjects. Often in electrodynamics, BCs were strongly associated with the context of E\&M waves and it can be worthwhile to shift the context to non-wave situations (e.g. going back to a static situtation).  In our physics class, students often only practice on simple situations (e.g. boundary z=0) and this may cause them to over-generalize the results to other situations. It is likely worthwhile to present students a variety of physical situations where there can be both simple and messier results. Students could also benefit from practice on recognizing unphysical limits and taking limits using appropriate independent variables to develop their reflection skills.

Application of the ACER framework provided an organizing structure for our analysis that helped us identify nodes in students' work where key difficulties appear. It also informed the development of interview protocols that targeted aspects of student problem solving not accessed by traditional exams. Other than the physical tool of BCs, the mathematical tool of complex exponentials is also essential to solve problems in the context of E\&M waves. For this paper, we limited the analysis to investigating student difficulties with BCs. Our ongoing research effort involves blending the ACER analyses of BC with complex exponentials and we will report our findings in future publications.

\begin{acknowledgments}
We gratefully acknowledge contributions of Andreas. Becker, Dimitri Dounas-Frazer and the PER{@}C group. This work was supported by CU Boulder, the CU Science Education Initiative and NSF-CCLI grant \#1023208.
\end{acknowledgments}

\bibliography{PERC-2015}

\end{document}